\documentclass[
 9pt,
 reprint,
 floats,
 aps,
 amssymb,
 floatfix,
 superscriptaddress]{revtex4-2}

\usepackage{subfiles}
\usepackage{subfigure}
\usepackage{graphicx}
\usepackage{color}
\usepackage{amsmath}
\usepackage{siunitx}
\usepackage{listings}
\usepackage[parfill]{parskip}
\usepackage{todonotes}

\graphicspath{{images/}{../images/}}

\begin{document}

\title{Machine learning accelerated computational fluid dynamics}

\author{Dmitrii Kochkov}
\thanks{D.K. and J.A. S. contributed equally to this work.}
\affiliation{Google Research}
\author{Jamie A. Smith}
\thanks{D.K. and J.A. S. contributed equally to this work.}
\affiliation{Google Research}
\author{Ayya Alieva}
\affiliation{Google Research}
\author{Qing Wang}
\affiliation{Google Research}
\author{Michael P. Brenner} 
\affiliation{Google Research}
\affiliation{School of Engineering and Applied Sciences, Harvard University, Cambridge, MA 02138}
\author{Stephan Hoyer}
\email{To whom correspondence should be addressed. E-mail: dkochkov@google.com, jamieas@google.com, shoyer@google.com}
\affiliation{Google Research}

\begin{abstract}
Numerical simulation of fluids plays an essential role in modeling many physical phenomena, such as weather, climate, aerodynamics and plasma physics.
Fluids are well described by the Navier-Stokes equations, but solving these equations at scale remains daunting, limited by the computational cost of resolving the smallest spatiotemporal features.
This leads to unfavorable trade-offs between accuracy and tractability.
Here we use end-to-end deep learning to improve approximations inside computational fluid dynamics for modeling two-dimensional turbulent flows.
For both direct numerical simulation of turbulence and large eddy simulation, 
our results are as accurate as baseline solvers with 8-10x finer resolution in each spatial dimension, resulting in 40-80x fold computational speedups.
Our method remains stable during long simulations, and generalizes to forcing functions and Reynolds numbers outside of the flows where it is trained, in contrast to black box machine learning approaches.
Our approach exemplifies how scientific computing can leverage machine learning and hardware accelerators to improve simulations without sacrificing accuracy or generalization.
\end{abstract}

\maketitle

\section{Introduction}

Simulation of complex physical systems described by nonlinear partial differential equations are  central to  engineering and physical science, with applications ranging from weather \cite{richardson2007weather,bauer2015} and climate \cite{schneider2017a,neumann2019} , engineering design of vehicles or engines \cite{anderson2009basic}, to wildfires \cite{bakhshaii2019review} and plasma physics~\cite{tang2005advances}.  Despite a direct link between the equations of motion and the basic laws of physics, it is impossible to carry out direct numerical simulations  at the scale required for these important problems. This fundamental issue has stymied progress in scientific computation for decades, and arises from the fact that an accurate simulation must resolve the smallest spatiotemporal scales.

A paradigmatic example is turbulent fluid flow \cite{pope2001turbulent}, underlying simulations of weather, climate, and aerodynamics. The size of the smallest eddy is tiny: for an airplane with chord length of 2 meters, the smallest length scale (the Kolomogorov scale) \cite{frisch1995turbulence} is $O(10^{-6})m$.
Classical methods for computational fluid dynamics (CFD), such as finite differences, finite volumes, finite elements and pseudo-spectral methods, are only accurate if fields vary smoothly on the mesh, and hence meshes must resolve the smallest features to guarantee convergence.
For a turbulent fluid flow,  the requirement to resolve the smallest flow features implies a computational cost scaling like ${\rm Re}^3$, where $Re= U L/\nu$, with $U$ and $L$ the typical velocity and length scales and $\nu$ the kinematic viscosity. A tenfold increase in ${\rm Re}$ leads to a thousandfold increase in the computational cost.  Consequently, direct numerical simulation (DNS) for e.g. climate and weather are impossible. Instead, it is traditional to use smoothed versions of the Navier Stokes equations \cite{moser2020statistical,meneveau2000scale} that allow coarser grids while sacrificing accuracy, such as Reynolds Averaged Navier-Stokes (RANS) \cite{Boussinesq1877-uf,Alfonsi2009-gt}, and Large-Eddy Simulation (LES) \cite{Smagorinsky1963-fj,Lesieur1996-zy}.
For example, current state-of-art LES with mesh sizes of $\mathcal{O}(10)$ to  $\mathcal{O}(100)$ million has been used in the design of internal combustion engines~\cite{Male2019-uk}, gas turbine engines~\cite{Wolf2012-fn,Esclapez2017-to}, and turbo-machinery~\cite{Arroyo2019-nq}. Despite promising progress in LES over the last two decades, there are severe limits to what can be accurately simulated. This is mainly due to the first-order dependence of LES on the sub-grid scale (SGS) model, especially for flows whose rate controlling scale is unresolved~\citep{Pope2004-ho}. 

\begin{figure}[t]
\centering
\includegraphics[width=\linewidth]{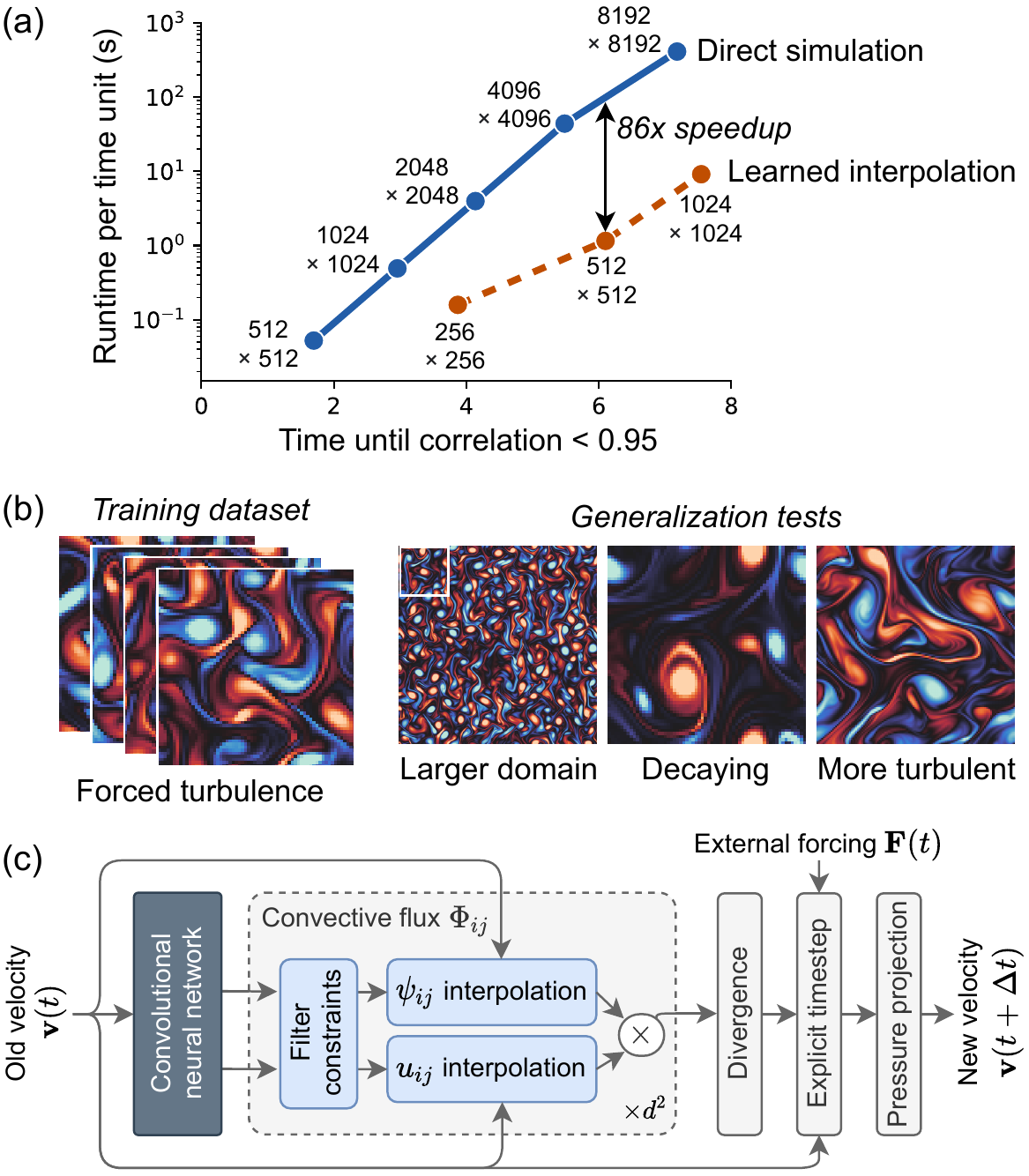}
\caption{Overview of our approach and results.
(a) Accuracy versus computational cost with our baseline (direct simulation) and ML accelerated (learned interpolation) solvers.
The $x$ axis corresponds to pointwise accuracy, showing how long the simulation is highly correlated with the ground truth, whereas the $y$-axis shows the computational time needed to carry out one simulation time-unit on a single TPU core.
Each point is annotated by the size of the corresponding spatial grid: for details see the appendix.
(b) Illustrative training and validation examples, showing the strong generalization capabilities of our model.
(c) Structure of a single time step for our "learned interpolation" model, with a convolutional neural net controlling learned approximations inside the convection calculation of a standard numerical solver.
$\psi$ and $u$ refer to advected and advecting velocity components.
For $d$ spatial dimensions there are $d^2$ replicates of the convective flux module, corresponding to the flux of each velocity component in each spatial direction.
}\label{fig:outline}
\end{figure}

Here, we introduce a method for calculating the accurate time evolution of solutions to nonlinear partial differential equations, while using an order of magnitude coarser grid than is traditionally required for the same accuracy.
This is a novel type of numerical solver that does not average unresolved degrees of freedom, but instead uses discrete equations that give pointwise accurate solutions on an unresolved grid.  We discover these algorithms using
machine learning, by replacing the components of traditional solvers most affected by the loss of resolution with learned alternatives.  
As shown in Fig.~\ref{fig:outline}(a), for a two dimensional direct numerical simulation of a turbulent flow, our algorithm maintains accuracy while using $10 \times$ coarser resolution in each dimension, resulting in a $\sim 80$ fold improvement in computational time with respect to an advanced numerical method of similar accuracy.  The model learns how to interpolate local features of solutions and hence can accurately  generalize to different flow conditions such as different forcings and even different Reynolds numbers [Fig.~\ref{fig:outline}(b)].  We also apply the method to a high resolution LES simulation of a turbulent flow and show similar performance enhancements, maintaining pointwise accuracy on $Re=100,000$ LES simulations using $8 \times$ times coarser grids with $\sim 40$ fold computational speedup. 

There has been a flurry of recent work using machine learning to improve turbulence modeling.
One major family of approaches uses ML to fit closures to classical turbulence models based on agreement with high resolution direct numerical simulations (DNS) \cite{Ling2016-qk,duraisamy2019,maulik2019subgrid,beck2019deep}.
While potentially more accurate than traditional turbulence models, these new models have not achieved reduced computational expense.
Another major thrust uses  “pure” ML, aiming to replace the entire Navier Stokes simulation with approximations based on deep neural networks \cite{kim2019deep,li2020neural,bhattacharya2020model,wang2020towards,lusch2018deep,erichson2019physics}.
A pure ML approach can be extremely efficient, avoiding the severe time-step constraints required for stability with traditional approaches.
Because these models do not include the underlying physics, they often struggle to enforce constraints, such as conservation of momentum and incompressibility.
While these models often perform well on data from the training distribution, they often struggle with generalization. For example, they perform worse when exposed to novel forcing terms.
A third approach, which we build upon in this work, uses ML to correct errors in cheap, under-resolved simulations \cite{sirignano2020dpm,um2020solver,pathak2020using}.
These models borrow strength from the coarse-grained simulations.

In this work we design algorithms that accurately solve the equations on coarser grids by replacing the components most affected by the  resolution loss with better performing learned alternatives.  We use {\sl data driven discretizations} \cite{bar2019learning,zhuang2020learned} to interpolate differential operators onto a coarse mesh with high accuracy [Fig.~\ref{fig:outline}(c)].
We train the solver inside a standard numerical method for solving the underlying PDEs as a differentiable program, with the neural networks and the numerical method  written in a framework (JAX~\cite{bradbury2020jax}) supporting reverse-mode automatic differentiation.
This allows for end-to-end gradient based optimization of the entire algorithm, similar to prior work on density functional theory~\cite{Li2020-kohn-sham-regularizer}, molecular dynamics~\cite{Schoenholz2020-jax-md} and fluids~\cite{sirignano2020dpm,um2020solver}.
The methods we derive are equation specific, and require training a coarse resolution solver with high resolution ground truth simulations. Since the dynamics of a partial differential equation are local, the high resolution simulations can be carried out on a small domain.  The models remains stable during long simulations and has robust and predictable generalization properties, with models trained on small domains producing accurate simulations on larger domains, with different forcing functions and even with different Reynolds number. Comparison to pure ML baselines shows that generalization arises from the physical constraints inherent in the formulation of the method.

\section{Background}\label{sec:background}

\subsection{Navier-Stokes}

Incompressible fluids are modeled by the Navier-Stokes equations:
\begin{subequations}
\label{eq:ns}
\begin{align}
    \label{eq:ns-momentum}
    \frac{\partial \mathbf{u}}{\partial t}
    =&
    - \nabla \cdot (\mathbf{u} \otimes \mathbf{u})
    + \frac{1}{\mathit{Re}} \nabla^2 \mathbf{u}
    - \frac{1}{\rho} \nabla p
    + \mathbf{f} \\
    \label{eq:ns-incompressible}
    \nabla \cdot \textbf{u} =& 0
\end{align}
\end{subequations}
where $\mathbf{u}$ is the velocity field, $\mathbf{f}$ the external forcing, and $\otimes$ denotes a tensor product.
The density $\rho$ is a constant, and the pressure $p$ is a Lagrange multiplier used to enforce \eqref{eq:ns-incompressible}.
The Reynolds number $\mathit{Re}$ dictates the balance between the convection (first) or diffusion (second) terms in the right hand side of \eqref{eq:ns-momentum}.
Higher Reynolds number flows dominated by convection are more complex and thus generally harder to model; flows are considered ``turbulent'' if $\mathit{Re} \gg 1$.

Direct numerical simulation (DNS) solves \eqref{eq:ns} directly, whereas large eddy simulation (LES) solves a spatially filtered version.
In the equations of LES, $\mathbf{u}$ is replaced by a filtered velocity $\overline{\mathbf{u}}$ and an sub-grid term $- \nabla \cdot \boldsymbol{\tau}$ is added to the right side of \eqref{eq:ns-momentum}, with the sub-grid stress defined as $\boldsymbol{\tau} = \overline{\mathbf{u} \otimes \mathbf{u}} - \overline{\mathbf{u}} \otimes \overline{\mathbf{u}}$.
Because $\overline{\mathbf{u} \otimes \mathbf{u}}$ is un-modeled, solving LES also requires a choice of \emph{closure} model for $\boldsymbol\tau$ as a function of $\overline{\mathbf{u}}$.
Numerical simulation of both DNS and LES further requires a \emph{discretization} step to approximate the continuous equations on a grid.
Traditional discretization methods (e.g., finite differences) converge to an exact solution as the grid spacing becomes small, with LES converging faster because it models a smoother quantity.
Together, discretization and closure models are the two principle sources of error when simulating fluids on coarse grids~\cite{sirignano2020dpm, Duraisamy2020-es}.

\begin{figure*}[!th]
\centering
\includegraphics[width=\linewidth]{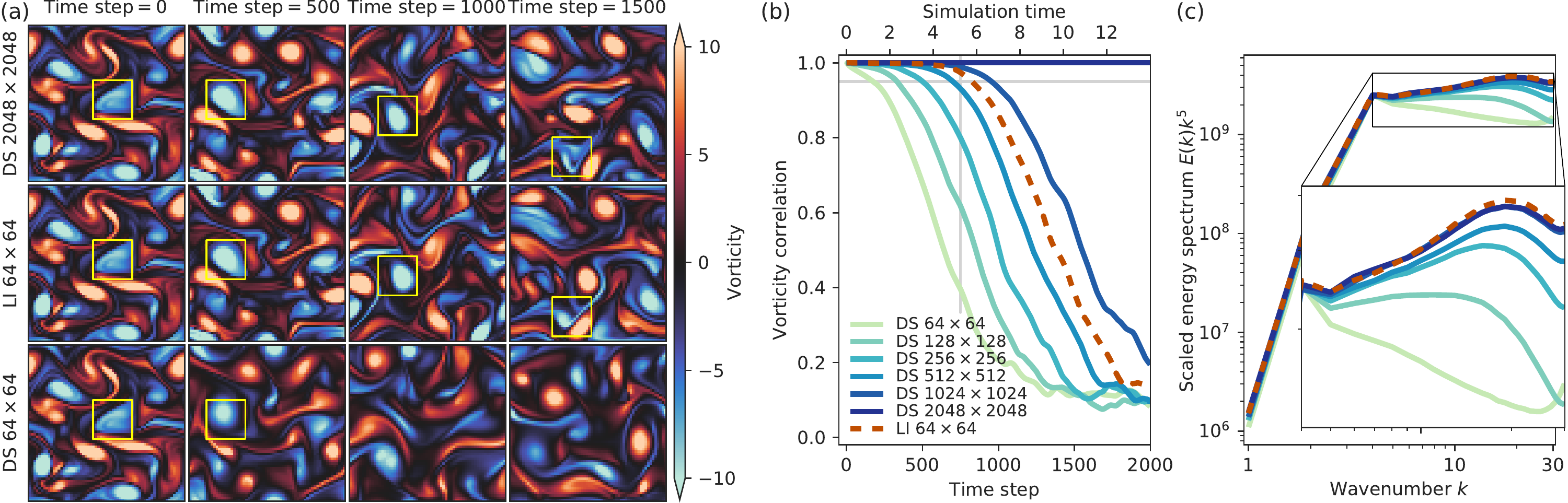}
\caption{Learned interpolation achieves accuracy of direct simulation at $\sim 10\times$ higher resolution. (a) Evolution of predicted vorticity fields for reference (DS $2048\times 2048$), learned (LI $64 \times 64$) and baseline (DS $64 \times 64$) solvers, starting from the same initial velocities. The yellow box traces the evolution of a single vortex. (b) Comparison of the vorticity correlation between predicted flows and the reference solution for our model and DNS solvers. (c) Energy spectrum scaled by $k^5$ averaged between time-steps 10000 and 20000, when all solutions have decorrelated with the reference solution.}\label{fig:accelerating_dns}
\end{figure*}

\subsection{Learned solvers}\label{subsec:challenges}

Our principle aim is to accelerate DNS without compromising accuracy or generalization.
To that end, we consider ML modeling approaches that enhance a standard CFD solver when run on inexpensive to simulate coarse grids.
We expect that ML models can improve the accuracy of the numerical solver via effective super-resolution of missing details~\cite{bar2019learning}.
Because we want to train neural networks for approximation \emph{inside} our solver, we wrote a new CFD code in JAX~\cite{bradbury2020jax}, which allows us to efficiently calculate gradients via automatic differentiation.
Our CFD code is a standard implementation of a finite volume method on a regular staggered mesh, with first-order explicit time-stepping for convection, diffusion and forcing, and implicit treatment of pressure; for details see the appendix.

The algorithm works as follows: in each time-step, the neural network generates a latent vector at each grid location based on the current velocity field, which is then used by the sub-components of the solver to account for local solution structure.
Our neural networks are convolutional, which enforces translation invariance and allows them to be local in space.
We then use components from standard standard numerical methods to enforce inductive biases corresponding to the physics of the Navier Stokes equations, as illustrated by the light gray boxes in Fig.~\ref{fig:outline}(c): the convective flux model improves the approximation of the discretized convection operator; the divergence operator enforces local conservation of momentum according to a finite volume method; the pressure projection enforces incompressibility and the explicit time step operator forces the dynamics to be continuous in time, allowing for the incorporation of additional time varying forces.
``DNS on a coarse grid'' blurs the boundaries of traditional DNS and LES modeling, and thus invites a variety of data-driven approaches.
In this work we focus on two types of ML components: learned interpolation and learned correction.
Both focus on the convection term, the key term in \eqref{eq:ns} for turbulent flows.

\subsubsection{Learned interpolation (LI)}

In a finite volume method, $\mathbf{u}$ denotes a vector field of volume averages over unit cells, and the cell-averaged divergence can be calculated via Gauss' theorem by summing the surface flux over the each face.
This suggests that our only required approximation is calculating the convective flux $\mathbf{u} \otimes \mathbf{u}$ on each face, which requires interpolating $\mathbf{u}$ from where it is defined.
Rather than using typical polynomial interpolation, which is suitable for interpolation without prior knowledge, here we use an approach that we call \emph{learned interpolation} based on data driven discretizations~\cite{bar2019learning}.
We use the outputs of the neural network to generate interpolation coefficients based on local features of the flow, similar to the fixed coefficients of polynomial interpolation.
This allows us to incorporate two important priors: (1) the equation maintains the same symmetries and scaling properties (e.g., rescaling coordinates ${\bf x}\to \lambda {\bf x}$) as the original equations, and (2) as the mesh spacing vanishes, the interpolation module retains polynomial accuracy so that our model performs well in the regime where traditional numerical methods excel.

\subsubsection{Learned correction (LC)}

An alternative approach, closer in spirit to LES modeling, is to simply model a residual correction to the discretized Navier-Stokes equations [\eqref{eq:ns}] on a coarse-grid~\cite{um2020solver, sirignano2020dpm}.
Such an approach generalizes traditional closure models for LES, but in principle can also account for discretization error.
We consider \emph{learned correction} models of the form $\mathbf{u}_{t} = \mathbf{u}^\ast_t + \text{LC}(\mathbf{u}^\ast_t)$, where LC is a neural network and $\mathbf{u}^\ast_t$ is the uncorrected velocity field from the numerical solver on a coarse grid.
Modeling the residual is appropriate both from the perspective of a temporally discretized closure model, and pragmatically because the relative error between $\mathbf{u}_t$ and $\mathbf{u}^\ast_t$ in a single time step is small.
LC models have fewer inductive biases than LI models, but they are simpler to implement and potentially more flexible.
We also explored LC models restricted to take the form of classical closure models (e.g., flow-dependent effective tensor viscosity models), but the restrictions hurt model performance and stability.

\subsubsection{Training} The training procedure tunes the machine learning components of the solver to minimize the discrepancy between an expensive high resolution simulation and a simulation produced by the model on a coarse grid. We accomplish this via supervised training where we use a cumulative point-wise error between the predicted and ground truth velocities as the loss function
$$L(x, y) = \sum\limits_{t_{i}}^{t_{T}} \textrm{MSE}(\textbf{u}(t_{i}), \tilde{\textbf{u}}(t_{i})).$$
The ground truth trajectories are obtained by using a high resolution simulation that is then coarsened to the simulation grid.
Including the numerical solver in the training loss ensures fully ``model consistent'' training where the model sees its own outputs as inputs \cite{Duraisamy2020-es, maulik2019subgrid, um2020solver}, unlike typical \emph{a priori} training where simulation is only performed offline.
As an example, for the Kolmogorov flow simulations below with Reynolds number $1000$, our ground truth simulation had a resolution of $2048$ cells along each spatial dimension. We subsample these ground truth trajectories along each dimension and time by a factor of $32$. For training we use $32$ trajectories of $4800$ sequential time steps each, starting from different random initial conditions. To evaluate the model, we generate much longer trajectories (tens of thousands of time steps) to verify that models remain stable and produce plausible outputs.
We unroll the model for 32 time steps when calculating the loss, which we find improves model performance for long time trajectories~\cite{um2020solver}, in some cases using gradient checkpoints at each model step to reduce memory usage~\cite{Griewank1994-yg}.

\begin{figure*}[!t]
\centering
\includegraphics[width=\linewidth]{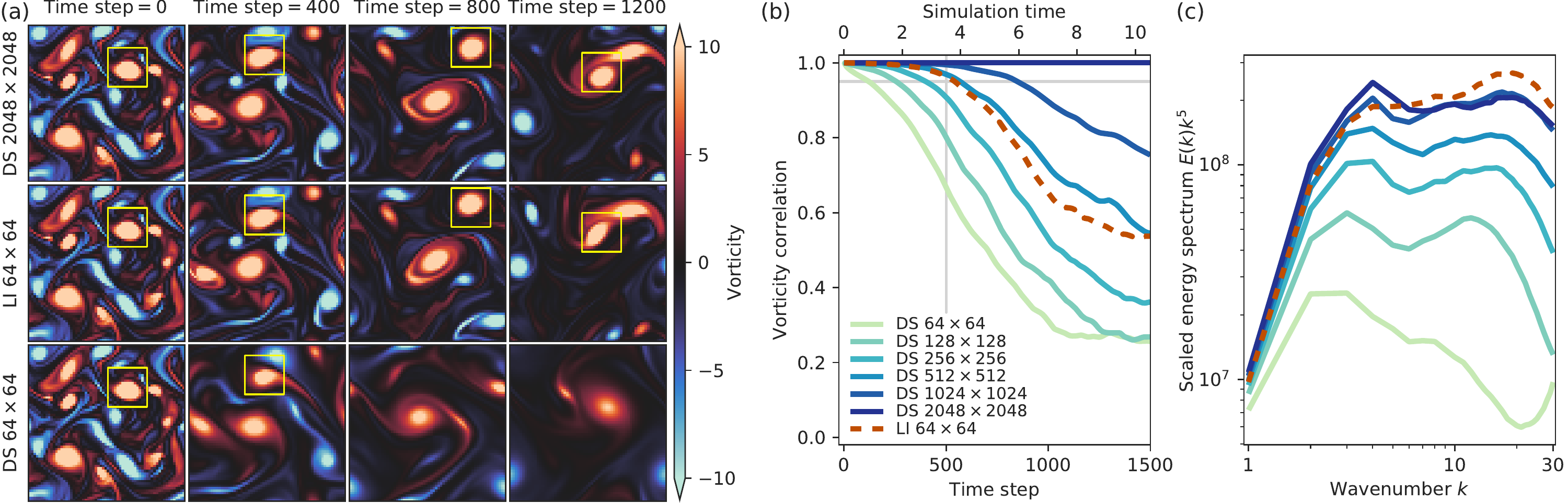}
\caption{Learned interpolation generalizes well to decaying turbulence. (a) Evolution of predicted vorticity fields as a function of time.
(b) Vorticity correlation between predicted flows and the reference solution.
(c) Energy spectrum scaled by $k^5$ averaged between time-steps 2000 and 2500, when all solutions have decorrelated with the reference solution.} \label{fig:decaying_generalization}
\end{figure*}

\section{Results}\label{sec:results}
We take a utilitarian perspective on model evaluation: simulation methods are good insofar as they demonstrate accuracy, computational efficiency and generalizability. 
In this case, accuracy and computational efficiency require the method to be faster than the DNS baseline, while maintaining accuracy for long term predictions; generalization means that although the model is trained on specific flows, it must be able to readily  generalize well to new simulation settings, including to different forcings and different Reynolds numbers.
In what follows, we first compare the accuracy and generalizability of our method to both direct numerical simulation and several existing ML-based approaches for simulations of two dimensional turbulence flow.
In particular, we first consider Kolmogorov flow~\cite{Chandler2013-koklmogorov}, a parametric family of forced turbulent flows  obeying the Navier-Stokes equation [\eqref{eq:ns}], with  forcing  $f = \sin(4y)\hat{\textbf{x}} - 0.1\textbf{u}$,
where the second term is a velocity dependent drag preventing accumulation of energy at large scales~\cite{Boffetta2012-2d-turbulence}.
Kolmogorov flow produces a statistically stationary turbulent flow, with flow complexity controlled by a single parameter, the Reynolds number $\textit{Re}$.

\subsection{Accelerating DNS} \label{subsec:accelerating_dns}
The accuracy of a direct numerical simulation quickly degrades once the grid resolution cannot capture the smallest details of the solution. In contrast, our ML-based approach strongly mitigates this effect. 
Figure \ref{fig:accelerating_dns} shows the results of training and evaluating our model on Kolmogorov flows at Reynolds number ${\rm Re}=1000$.  All datasets were generated using high resolution DNS, followed by a coarsening step. 

\subsubsection{Accuracy}\label{subsubsec:accelerating_dns_performance}

The scalar vorticity field $\omega = \partial_x u_y - \partial_y u_x$ is a convenient way to describe a two-dimensional incompressible flows~\cite{Boffetta2012-2d-turbulence}.
Accuracy can be quantified by correlating vorticity fields,\footnote{In our case the Pearson correlation reduces to a cosine distance because the flows considered here have mean velocity of $0$.} $C(\omega, \hat\omega)$ between the ground truth solution $\omega$ and the predicted state $\hat\omega$.
Fig.~\ref{fig:accelerating_dns} compares the learned interpolation model ($64 \times 64$) to fully resolved DNS of Kolmogorov flow  ($2048 \times 2048$) using an initial condition that was not included in the training set.
Strikingly, the learned discretization model matches the pointwise accuracy of DNS with a $\sim10$ times finer grid.
The eventual loss of correlation with the reference solution is expected due to the chaotic nature of turbulent flows; this is marked by a vertical grey line in Fig.~\ref{fig:accelerating_dns}(b), indicating the first three Lyapunov times. Fig.~\ref{fig:accelerating_dns} (a) shows the time evolution of the vorticity field for three different models: the learned interpolation matches the ground truth ($2048 \times 2048$) more accurately than the $512 \times 512$ baseline, whereas it greatly outperforms a baseline solver at the  same resolution as the model ($64 \times 64$).

The learned interpolation model also produces a similar energy spectrum $E(\textbf{k})=\frac{1}{2} |\textbf{u}(\textbf{k})|^{2}$ to DNS.
With decreasing resolution, DNS cannot capture high frequency features, resulting in an energy spectrum that ``tails off'' for higher values of $k$. 
Fig. \ref{fig:accelerating_dns} (c) compares the energy spectrum for learned interpolation and direct simulation at different resolutions after $10^4$ time steps.
The learned interpolation model accurately captures the energy distribution across the spectrum.

\subsubsection{Computational efficiency}

The ability to match DNS with a $\sim 10$ times coarser grid makes the learned interpolation solver much faster.
We benchmark our solver on a single core of Google's Cloud TPU v4, a hardware accelerator designed for accelerating machine learning models that is also suitable for many scientific computing use-cases~
\cite{Ben-Haim2019-ja, Yang2019-si, Lu2020-cb}.
The TPU is designed for high throughput vectorized operations, and extremely high throughput matrix-matrix multiplication in low precision (bfloat16).
On sufficiently large grid sizes ($256 \times 256$ and larger), our neural net makes good use of matrix-multiplication unit, achieving 12.5x higher throughput in floating point operations per second than our baseline CFD solver.
Thus despite using 150 times more arithmetic operations, the ML solver is only about 12 times slower than the traditional solver at the same resolution.
The $10 \times$ gain in effective resolution in three dimensions (two space dimensions and time, due to the Courant condition) thus corresponds to a speedup of $10^3/12 \approx 80$.

\begin{figure*}[t]
\centering
\includegraphics[width=\linewidth]{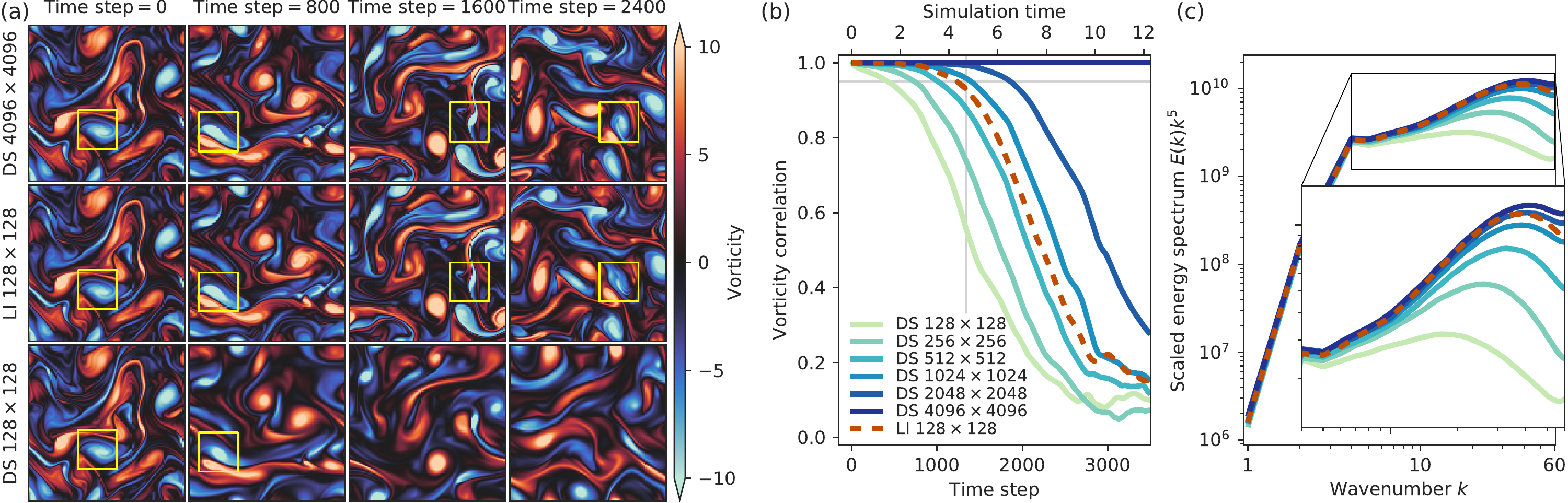}
\caption{Learned interpolations can be scaled to simulate higher Reynolds numbers without retraining.
(a) Evolution of predicted vorticity fields as a function of time for Kolmogorov flow at $Re=4000$.
(b) Vorticity correlation between predicted flows and the reference solution.
(c) Energy spectrum scaled by $k^5$ averaged between time-steps 6000 and 12000, when all solutions have decorrelated with the reference solution.
}\label{fig:kolmogorov_re_4000}
\end{figure*}

\subsubsection{Generalization}\label{subsubsec:accelerating_dns_generalization} 
In order to be useful, a learned model must accurately simulate flows outside of the training distribution.
We expect our models to generalize well because they learn local operators: interpolated values and corrections at a given point depend only on the flow within a small neighborhood around it.
As a result, these operators can be applied to any flow that features similar local structures as those seen during training.
We consider three different types of generalization tests: (1) larger domain size, (2) unforced decaying turbulent flow, and (3) Kolmogorov flow at a larger Reynolds number.

First, we test generalization to larger domain sizes with the same forcing. 
Our ML models have essentially the exact same performance as on the training domain, because they only rely upon local features of the flows.(see Appendix \ref{apx:pointwise_accuracy_extra} and Fig.~\ref{fig:model_comparison}). 

Second, we apply our model trained on Kolmogorov flow to {\sl decaying turbulence}, by starting with a random initial condition with high wavenumber components, and letting the turbulence evolve in time without forcing.
Over time, the small scales coalesce to form large scale structures, so that both the scale of the eddies and the Reynolds number vary.
Figure \ref{fig:decaying_generalization} shows that a learned discretization model trained on Kolmogorov flows ${\rm Re=1000}$ can match the accuracy of DNS running at $\sim 7$ times finer resolution.
A standard numerical method at the same resolution as the learned discretization model is corrupted by numerical diffusion, degrading the energy spectrum as well as pointwise accuracy.

Our final generalization test is harder: can the models generalize to higher Reynolds number where the flows are more complex? 
The universality of the turbulent cascade \cite{richardson2007weather,kolmogorov1941local,kraichnan1980two} implies that at the size of the smallest eddies (the Kolmogorov length scale), flows ``look the same'' regardless of Reynolds number when suitably rescaled.
This suggests that we can apply the model trained at one Reynolds number to a flow at another Reynolds number by simply rescaling the model to match the new smallest length scale. 
To test this we construct a new dataset for a Kolmogorov flow with ${\rm Re}=4000$. The theory of two-dimensional turbulence \cite{boffetta2012two} implies that the smallest eddy size decreases as $1 / \sqrt{\rm Re}$, implying that the smallest eddies in this flow are $1/2$ that for original flow with  ${\rm Re}=1000$.
We therefore can use a trained ${\rm Re}=1000$ model at ${\rm Re}= 4000$ by simply halving the grid spacing.
Fig. \ref{fig:kolmogorov_re_4000} (a) shows that with this scaling,  our model achieves the accuracy of DNS running at $7$ times finer resolution. This degree of generalization is remarkable, given that we are now testing the model with a flow of substantially greater complexity. Fig. \ref{fig:kolmogorov_re_4000} (b)  visualizes the vorticity, showing that higher complexity is captured correctly, as is further verified by the energy spectrum shown in Fig. \ref{fig:kolmogorov_re_4000} (c).

\subsection{Comparison to other ML models}\label{subsec:model_comparison}
Finally, we compare the performance of learned interpolation to alternative ML-based methods.
We consider three popular ML methods: ResNet (RN)~\cite{he2016deep}, Encoder-Processor-Decoder (EPD)~\cite{battaglia2018relational,sanchez2020learning} architectures and the learned correction (LC) model introduced earlier.
These models all perform explicit time-stepping without any additional latent state beyond the velocity field, which allows them to be evaluated with arbitrary forcings and boundary conditions, and to use the time-step based on the CFL condition.
By construction, these models are inivariant to translation in space and time, and have similar runtime for inference (varying within a factor of two).
To evaluate training consistency, each model is trained 9 times with different random initializations on the same Kolmogorov ${\rm Re}=1000$ dataset described previously.
Hyperparameters for each model were chosen as detailed in Appendix \ref{apx:hyperparameters}, and the models are evaluated on the same generalization tasks.
We compare their performance using several metrics: time until vorticity correlation falls below $0.95$ to measure pointwise accuracy for the flow over short time windows, the absolute error of the energy spectrum scaled by $k^{5}$ to measure statistical accuracy for the flow over long time windows,  and the fraction of simulated velocity values that does not exceed the range of the training data to measure stability.

\begin{figure}
\centering
\includegraphics[width=\linewidth]{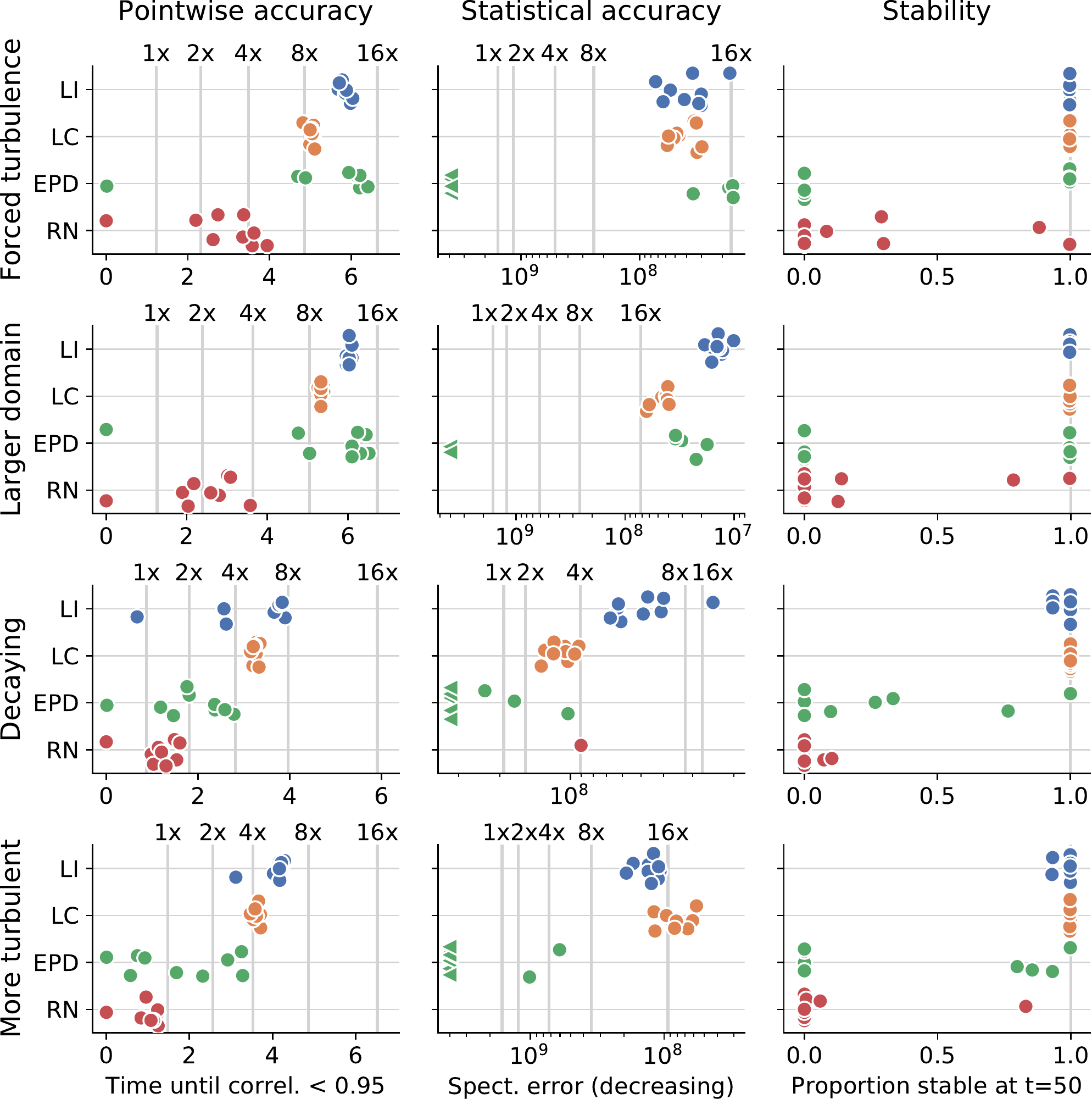}
\caption{Learned discretizations outperform a wide range of baseline methods in terms of accuracy, stability and generalization.
Each row within a subplot shows performance metrics for nine model replicates with the same architecture, but different randomly initialized weights.
The models are trained on forced turbulence, with larger domain, decaying and more turbulent flows as generalization tests. 
Vertical lines indicate performance of non-learned baseline models at different resolutions (all baseline models are perfectly stable).
The pointwise accuracy test measures error at the start of time integration, whereas the statistical accuracy and stability tests are both performed on simulations at time 50 after about $7000$ time integration steps (twice the number of steps for more turbulent flow).
The points indicated by a left-pointing triangle in the statistical accuracy tests are clipped at a maximum error.}
\label{fig:model_comparison}
\end{figure}

Fig.~\ref{fig:model_comparison} compares  results across all considered configurations.
Overall, we find that learned interpolation (LI) performs best, although learned correction (LC) is not far behind.
We were impressed by the performance of the LC model, despite its weaker inductive biases.
The difference in effective resolution for pointwise accuracy ($8 \times$ vs $10 \times$ upscaling) corresponds to about a factor of two in run-time.
There are a few isolated exceptions where pure black box methods outperform the others, but not consistently.
A particular strength of the learned interpolation and correction models is their consistent performance and generalization to other flows, as seen from the narrow spread of model performance for different random initialization and their consistent dominance over other models in the generalization tests.
Note that even a modest $4 \times$ effective coarse-graining in resolution still corresponds to a $5 \times$ computational speed-up.
In contrast, the black box ML methods exhibit high sensitivity to random initialization and do not generalize well, with much less consistent statistical accuracy and stability.

\subsection{Acceleration of LES}

Finally, up until now we have illustrated our method for DNS of the Navier Stokes equations.
Our approach is quite general and could be applied to any nonlinear partial differential equation.
To demonstrate this, we apply the method to accelerate LES, the industry standard method for large scale simulations where DNS is not feasible.

\begin{figure*}
\centering
\includegraphics[width=\linewidth]{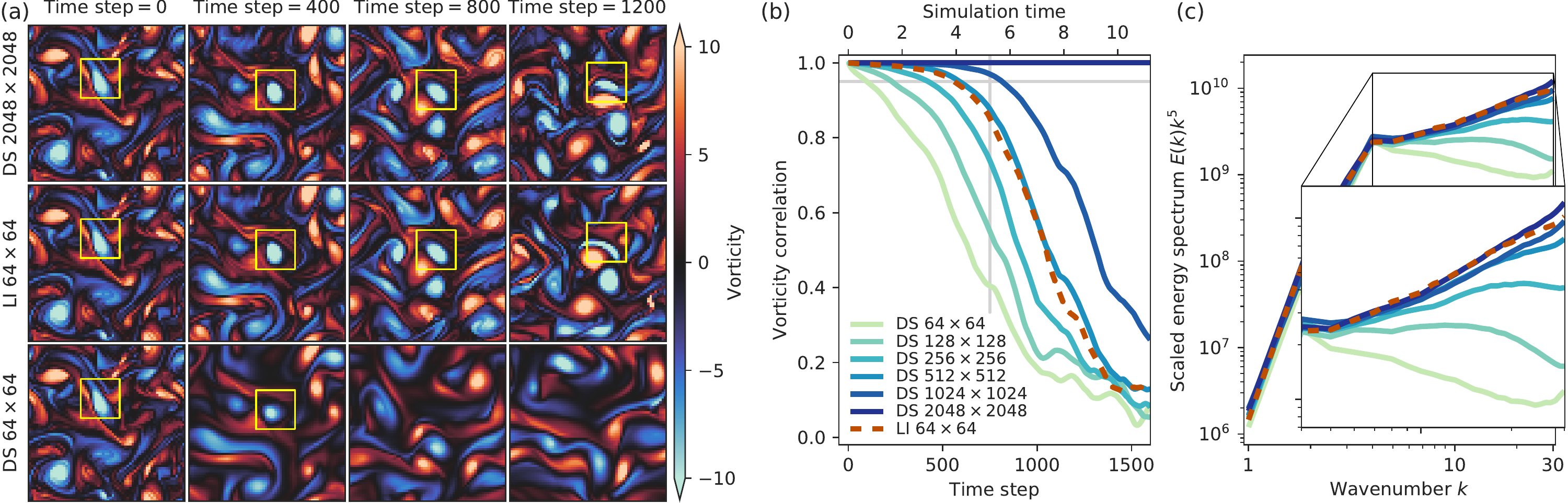}
\caption{Learned discretizations achieve accuracy of LES simulation running on $8 \times$ times finer resolution.
(a) Evolution of predicted vorticity fields as a function of time.
(b) Vorticity correlation between predicted flows and the reference solution.
(c) Energy spectrum scaled by $k^5$ averaged between time-steps 3800 and 4800, when all solutions have decorrelated with the reference solution.
}\label{fig:accelerating_les}
\end{figure*}

Here we treat the LES at high resolution as the ground truth simulation and train an interpolation model on a coarser grid for Kolmogorov flows with Reynolds number $\textit{Re}=10^5$ according to the Smagorinsky-Lilly model SGS model~\cite{pope2001turbulent}.
Our training procedure follows the exact same approach we used for modeling DNS.
Note in particular that we do not attempt to model the parameterized viscosity in the learned LES model, but rather let learned interpolation model this implicitly.
Fig.~\ref{fig:accelerating_les} shows that learned interpolation for LES still achieves an effective $8 \times$ upscaling, corresponding to roughly $40 \times$ speedup.

\section{Discussion }

In this work we present a data driven numerical method that achieves the same accuracy as traditional finite difference/finite volume methods but with much coarser resolution.
The method learns accurate local operators for convective fluxes and residual terms, and  matches the accuracy of an advanced numerical solver running at 8--$10\times$ finer resolution, while performing the computation 40--$80\times$ faster.  The method uses machine learning to  interpolate better at a coarse scale, within the framework of the traditional numerical discretizations. As such, the method inherently contains the scaling and symmetry properties of the original governing Navier Stokes equations. For that reason, the methods generalize much better than pure black-box machine learned methods, not only to different forcing functions but also to different parameter regimes (Reynolds numbers).  

What outlook do our results suggest for speeding up 3D turbulence? 
In general, the runtime $T$ for efficient ML augmented simulation of time-dependent PDEs should scale like
\begin{equation}
    T \sim \left( C_\text{ML} + C_\text{physics} \right) \left( \frac{N}{K} \right)^{d+1},
\end{equation}
where $C_\text{ML}$ is the cost of ML inference per grid point, $C_\text{physics}$ is the cost of baseline numerical method, $N$ is the number of grid points along each dimension of the resolved grid, $d$ is the number of spatial dimensions and $K$ is the effective coarse graining factor.
Currently, $C_\text{ML} / C_\text{physics} \approx 12$, but we expect that much more efficient machine learning models are possible, e.g., by sharing work between time-steps with recurrent neural nets.
We expect the $10 \times$ decrease in effective resolution discovered here to generalize to 3D and more complex problems. This suggests that speed-ups in the range of $10^3$--$10^4$ may be possible for 3D simulations.
Further speed-ups, as required to capture the full range of turbulent flows, will require either more efficient representations for flows (e.g., based on solution manifolds rather than a grid) or being satisfied with statistical rather than pointwise accuracy (e.g., as done in LES modeling).

In summary, our approach expands the Pareto frontier of efficient simulation in computational fluid dynamics, as illustrated in Fig.~\ref{fig:outline}(a).
With ML accelerated CFD, users may either solve expensive simulations much faster, or increase accuracy without additional costs.
To put these results in context, if applied to numerical weather prediction, increasing the duration of accurate predictions from 4 to 7 time-units would correspond to approximately 30 years of progress~\cite{Bauer2015-quiet-rev-nwp}.
These improvements are possible due to the combined effect of two technologies still undergoing rapid improvements:
modern deep learning models, which allow for accurate simulation with much more compact representations,
and modern accelerator hardware, which allows for evaluating said models with a remarkably small increase in computational cost.
We expect both trends to continue for the foreseeable future, and to eventually impact all areas of computationally limited science.

\section*{Acknowledgement}
We thank John Platt and Rif A.\ Saurous for encouraging and supporting this work and for important conversations, and Yohai bar Sinai, Anton Geraschenko, Yi-fan Chen and Jiawei Zhuang for important conversations.

\appendix
\renewcommand{\thefigure}{A\arabic{figure}}
\setcounter{figure}{0}
\renewcommand{\thetable}{A\arabic{table}}
\setcounter{table}{0}
\section{Direct numerical simulation}\label{apx:dns}
Here we describe the details of the numerical solver that we use for data generation, model comparison and the starting point of our machine learning models. 
Our solver uses a staggered-square mesh \cite{mcdonough2007lectures} in a finite volume formulation: the computational domain is broken into computational cells where the velocity field is placed on the edges, while the pressure is solved at the cell centers. Our choice of real-space formulation of the Navier-Stokes equations, rather than a spectral method is motivated by practical considerations: real space simulations are much more versatile when dealing with boundary conditions and non-rectangular geometries. We now describe the implementation details of each component.

\subsection*{Convection and diffusion}
We implement convection and diffusion operators based on finite-difference approximations. The Laplace operator in the diffusion is approximated using a second order central difference approximation. The convection term is solved by advecting all velocity components simultaneously, using a high order scheme based on Van-Leer flux limiter \cite{sweby1984high}. For the results presented in the paper we used explicit time integration using Euler discretization. This choice is motivated by performance considerations: for the simulation parameters used in the paper (high Reynold number) implicit diffusion is not required for stable time-stepping, and is approximately twice as slow, due to the additional linear solves required for each velocity component. For diffusion dominated simulations implicit diffusion would result in faster simulations.

\subsection*{Pressure}
To account for pressure we use a projection method, where at each step we solve the corresponding Poisson equation. The solution is obtained using either a fast diagonalization approach with explicit matrix multiplication \cite{Lynch1964-pe} or a real-valued fast Fourier transform (FFT). The former is well suited for small simulation domains as it has better accelerator utilization, while FFT has best computational complexity and performs best in large simulations. For wall-clock evaluations we choose between the fast diagonalization and FFT approach by choosing the fastest for a given grid.

\subsection*{Forcing and closure terms}
We incorporate forcing terms together with the accelerations due to convective and diffusive processes. In an LES setting the baseline and ground truth simulations additionally include a subgrid scale model that is also treated explicitly. We use the Smagorinsky-Lilly model (\eqref{apx:eq:les_model}) where $\Delta$ is the grid spacing and $C_{s}=0.2$:

\begin{align} \label{apx:eq:les_model}
    \tau_{i j}=-2\left(C_{s} \Delta\right)^{2}&|\bar{S}| \bar{S}_{i j} \\
    \bar{S}_{i j}= \frac{1}{2}(\partial_{i}u_{j} + \partial_{j}u_{i}) \qquad &|\bar{S}| = 2\sqrt{\sum\limits_{i, j} \bar{S}_{i j} \bar{S}_{i j}} \nonumber
\end{align}

\section{Datasets and simulation parameters}\label{apx:simulation_setup}
In the main text we introduced five datasets: two Kolmogorov flows at $\textit{Re} = 1000$ and $\textit{Re} = 4000$, Kolmogorov flow with $\textit{Re} = 1000$ on a $2 \times$ larger domain, decaying turbulence and an LES dataset with Reynolds number $10^5$. Dataset generation consisted of three steps: (1) burn-in simulation from a random initial condition; (2) simulation for a fixed duration using high resolution solver; (3) downsampling of the solution to a lower grid for training and evaluation.

The burn-in stage is fully characterized by the \textit{burn-in time} and \textit{initialization wavenumber} which represent the discarded initial transient and the peak wavenumber of the log-normal distribution from which random initial conditions were sampled from. The maximum amplitude of the initial velocity field was set to $7$ for forced simulations and $4.2$ in the decaying turbulence, which was selected to minimize the burn-in time, as well as maintain standard deviation of the velocity field close to $1.0$. The initialization wavenumber was set to $4$. Simulation parameters include \textit{simulation resolution} along each spatial dimension, \textit{forcing} and \textit{Reynolds number}. The resulting velocity trajectories are then downsampled to the \textit{save resolution}. Note that besides the spatial downsampling we also perform downsampling in time to maintain the Courant–Friedrichs–Lewy (CFL) factor fixed at $0.5$, standard for numerical simulations. Parameters specifying all five datasets are shown in Table \ref{table:simulation_params}. We varied the size of our datasets from $12200$ time slices at the downsampled resolution for decaying turbulence to $34770$ slices in the forced turbulence settings. Such extended trajectories allow us to analyze stability of models when performing long-term simulations. We note that when performing simulations at higher Reynolds numbers we used rescaled grids to maintain fixed time stepping. This is motivated to potentially allow methods to specialize on a discrete time advancements. When reporting the results, spatial and time dimensions are scaled back to a fixed value to enable direct comparisons across the experiments.

For comparison of our models with numerical methods we use the corresponding simulation parameters while changing only the resolution of the underlying grid. As mentioned in the Appendix \ref{apx:dns}, when measuring the performance of all solvers on a given resolution we choose solver components to maximize efficiency.

\begin{table}[]
\begin{tabular}{|l|l|l|l|}
\hline
\textbf{Dataset}       & \textbf{Resolution}               & \textbf{Re}              & \textbf{Burn-in time} \\ \hline
Kolmogorov $\textit{Re} = 1000$ & $2048 \to 64$                     & $1000$                   & $40$                  \\ \hline
Kolmogorov $\textit{Re} = 4000$ & $4096 \to 128$                    & $4000$                   & $40$                  \\ \hline
Kolmogorov $\textit{Re} = 1000$ & $4096 \to 128$                     & $1000$                   & $40$                  \\ \hline
Decaying turbulence    & $2048 \to 64$                     & $1000$ ($t=0$)          & $4$                   \\ \hline
LES $\textit{Re}=10^5$        & $2048 \to 64$                     & $10^5$                 & $40$                  \\ \hline
\end{tabular}
\caption{Simulation parameters used for generation of the datasets in this manuscript. For resolution, the notation $X \to Y$ indicates that simulation was performed at resolution $X$ and the result was downsampled to resolution $Y$.}\label{table:simulation_params}
\end{table}

\section{Learned interpolations}\label{apx:learned_interpolations}
To improve numerical accuracy of the convective process our models use psuedo-linear models for interpolation~\cite{bar2019learning, zhuang2020learned}.
The process of interpolating $u_{i}$ to $u(x)$ is broken into two steps:
\begin{enumerate}
    \item Computation of local stencils for interpolation target $x$.
    \item Computation of a weighted sum $\sum_{i=1}^n a_{i} u_{i}$ over the stencil.
\end{enumerate}
Rather using a fixed set of interpolating coefficients $a_i$ (e.g., as done for typical polynomial interpolation), we choose $a_i$ from the output of a neural network additionally constrained to satisfy $\sum_{i=1}^n a_i = 1$, which guarantees that the interpolation is at least first order accurate.
We do so by generating interpolation coefficients with an affine transformation $\mathbf{a} = A\mathbf{x} + \mathbf{b}$ on the unconstrainted outputs $x$ of the neural network, where $A$ is the null-space of the constraint matrix (a $1 \times n$ matrix of ones) of shape $n \times (n - 1)$ and $\mathbf{b}$ is an arbitrary valid set of coefficients (we use linear interpolation from the nearest source term locations).
This is a special case of the procedure for arbitrary polynomial accuracy constraints on finite difference coefficients described in \cite{bar2019learning, zhuang2020learned}.
In this work, we use a $4 \times 4$ patch centered over the top-right corner of each unit-cell, which means we need $15$ unconstrained neural network outputs to generate each set of interpolation coefficients.

\section{Neural network architectures}

\begin{figure}
    \centering
    \includegraphics[width=\linewidth]{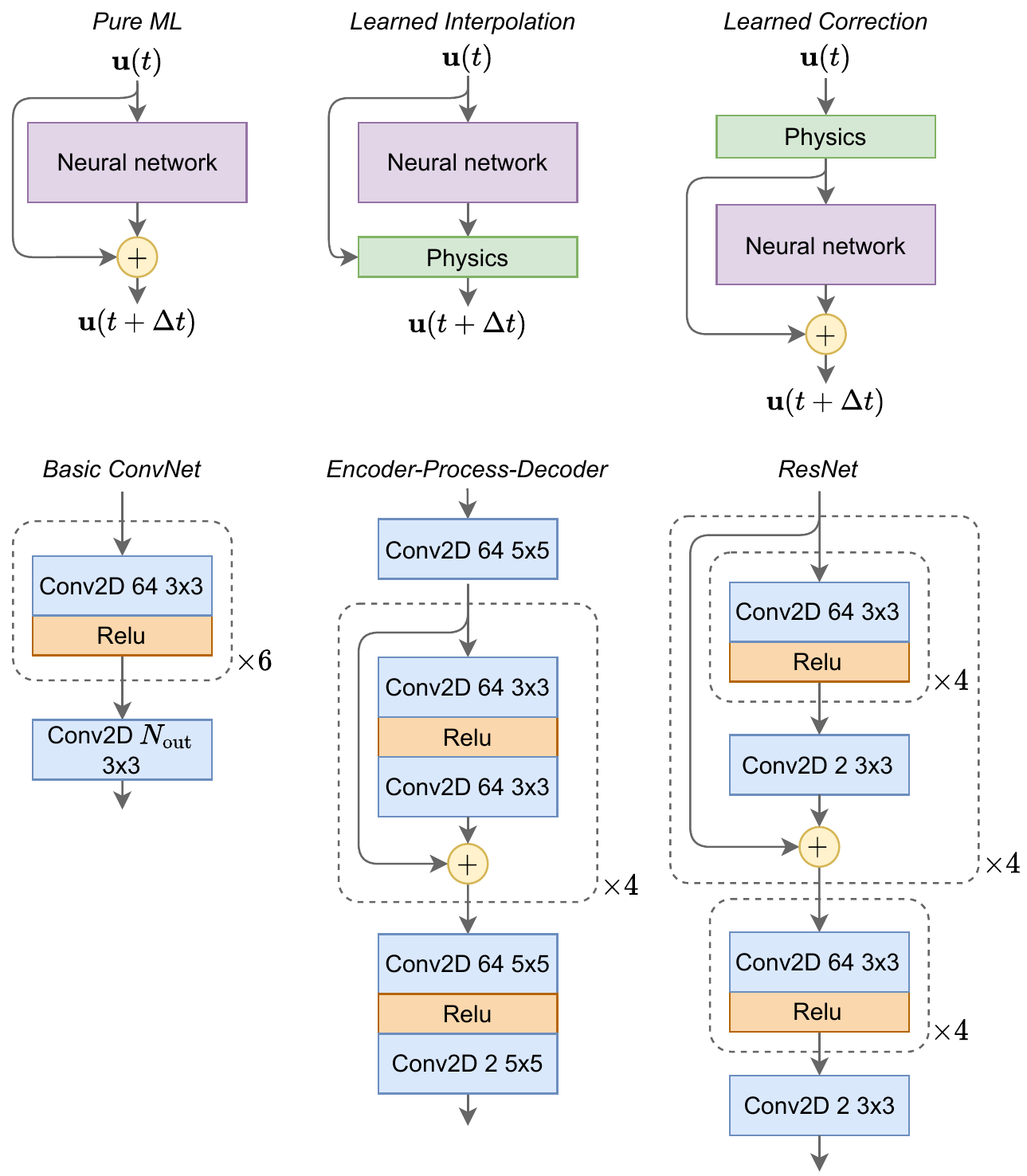}
    \caption{Modeling approaches and neural network architecutres used in this paper. Rescaling of inputs and outputs from neural network sub-modules with fixed constants is not depicted. Dashed lines surround components that are repeatedly applied the indicated number of times.}
    \label{fig:model_architectures}
\end{figure}

All of the ML based models used in this work are based on fully convolutional architectures. Fig.~\ref{fig:model_architectures} depicts our three modeling approaches (pure ML, learned interpolation and learned correction) and three architecturs for neural network sub-components (Basic ConvNet, Encoder-Process-Decoder and ResNet).
Outputs and inputs for each neural network layer are linearly scaled such that appropriate values are in the range $(-1, 1)$.

For our physics augmented solvers (learned interpolation and learned correction), we used the basic ConvNet archicture, with $N_\text{out} = 120$ (8 interpolations that need 15 inputs each) for learned interpolation and $N_\text{out} = 2$ for learned correction.
Our experiments found accuracy was slightly improved by using larger neural networks, but not sufficiently to justify the increased computational cost.

For pure ML solvers, we used EPD and ResNet models that do not build impose physical priors beyond time continuity of the governing equations.
In both cases a neural network is used to predict the acceleration due to physical processes that is then summed with the forcing function to compute the state at the next time.
Both EPD and ResNet models consist of a sequence of CNN blocks with skip-connections.
The main difference is that the EPD model has an arbitrarily large hidden state (in our case, size 64), whereas the ResNet model has a fixed size hidden state equal to 2, the number of velocity components.

\section{Details of accuracy measurements}\label{apx:pointwise_accuracy_extra}
\begin{figure}[t]
    \centering
    \includegraphics[width=\linewidth]{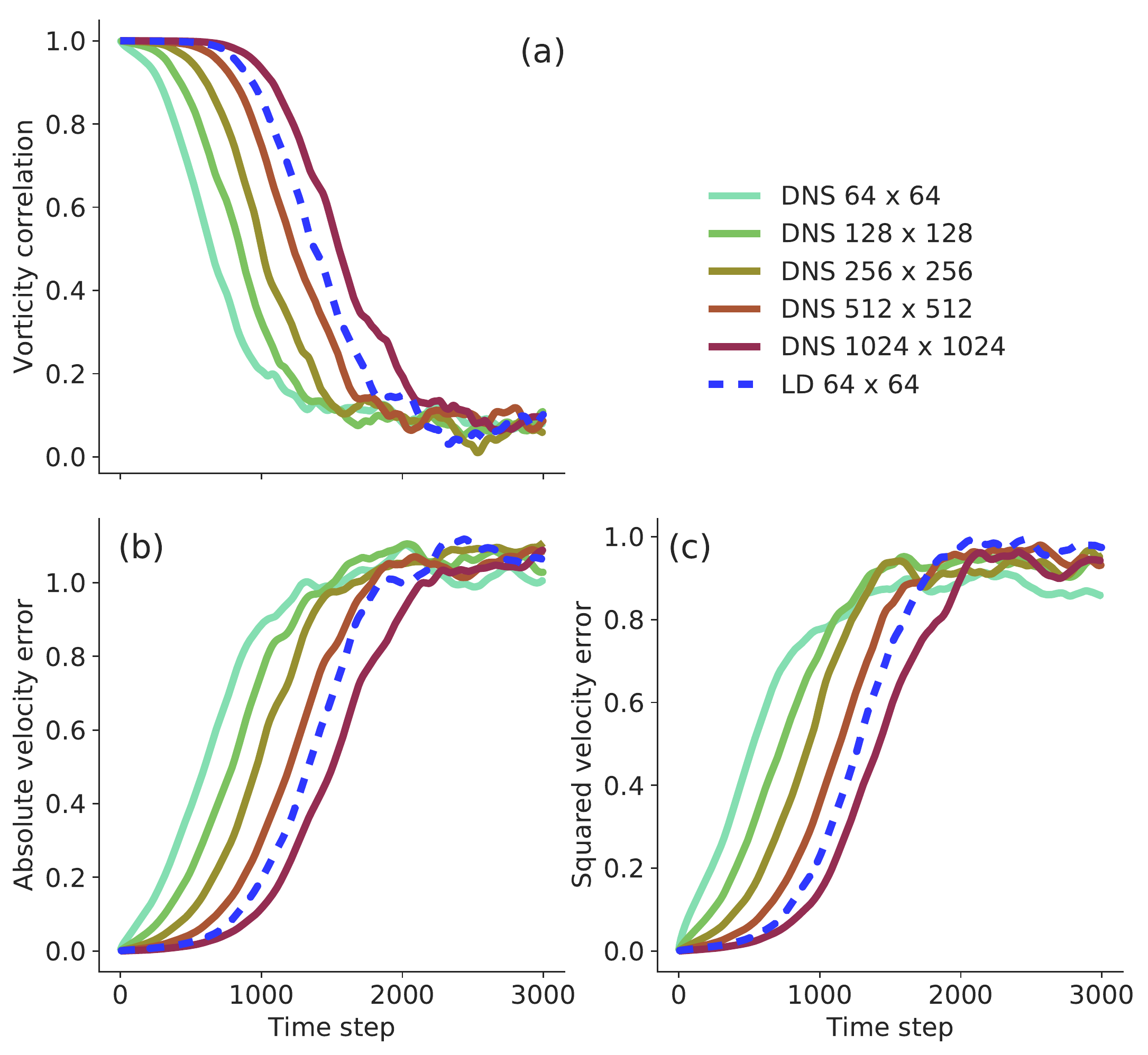}
    \caption{Comparison of ML + CFD model to DNS running at different resolutions using varying metrics. (a) Vorticity correlation as presented in the main text. (b) Mean absolute error. (c) Absolute error of the kinetic energy.}
    \label{fig:mae_mse_errors}
\end{figure}

In the main text we have presented accuracy results based on correlation between the predicted flow configurations and the reference solution. We reach the same conclusions based on other common choices, such as squared and absolute errors as shown in Fig. \ref{fig:mae_mse_errors} comparing learned discretizations to DNS method on Kolmogorov flow with Reynolds number $1000$.

\begin{figure*}
    \centering
    \includegraphics[width=\linewidth]{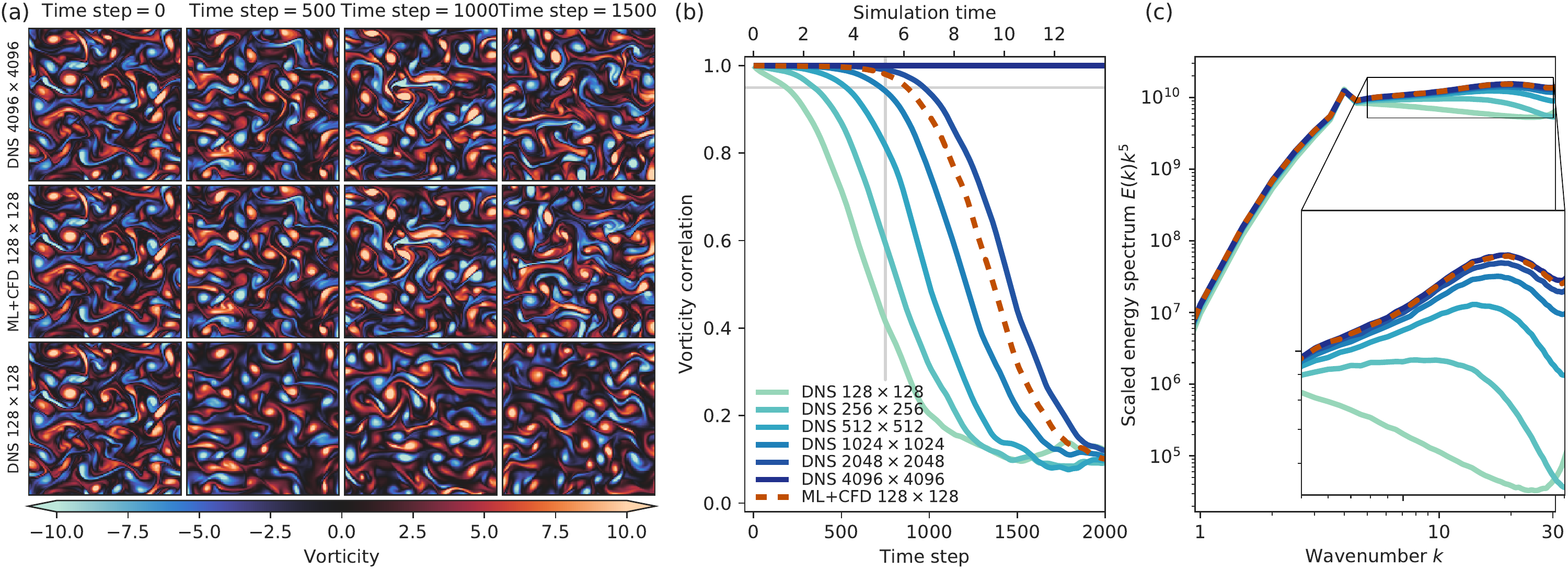}
    \caption{Comparison of ML + CFD model to DNS running at different resolutions when evaluated on a $2 \times$ larger domain with characteristic length-scales matching those of the training data. (a) Visualization of the vorticity at different times. (b) Vorticity correlation as a function of time. (c) Scaled energy spectrum $E(k) * k^{5}$.}
    \label{fig:larger_domain_generalization}
\end{figure*}

As mentioned in the main text, we additionally evaluated our models on enlarged domains while keeping the flow complexity fixed.
Because our models are local, this is the simplest generalization test.
As shown in Fig.~\ref{fig:larger_domain_generalization} (and Fig.~\ref{fig:model_comparison} in the main text), the improvement for $2 \times$ larger domains  is identical to that found on a smaller domain.

\section{Details of overview figure}

The Pareto frontier of model performance shown in Fig.~\ref{fig:outline}(a) is based on extrapolated accuracy to an $8 \times$ larger domain.
Due to computational limits, we were unable to run the simulations on the $16384 \times 16384$ grid for measuring accuracy, so the time until correlation goes below $0.95$ was instead taken from the $1 \times$ domain size, which as described in the preceding section matched performance on the $2 \times$ domain.

The $512 \times 512$ learned interpolation model corresponds to the depicted results in Fig.~\ref{fig:accelerating_dns}, whereas the $256 \times 256$ and $1024 \times 1024$ models were retrained on the same training dataset for $2 \times$ coarser or $2 \times$ finer coarse-graining.

\section{Hyperparameter tuning}\label{apx:hyperparameters}
All models were trained using Adam \cite{kingma2014adam} optimizer. Our initial experiments showed that none of the models had a consistent dependence on the optimization parameters. The set that worked well for all models included learning rate of $10^{-3}$, $b1=0.9$ and $b2=0.99$. For each model we performed a hyperparameter sweep over the length of the training trajectory $t_{T}$ used to compute the loss, model capacity such as size and number of hidden layers, as well as a sweep over different random initializations to assess reliability and consistency of training.

\bibliography{main}

\end{document}